\documentclass[aps,prb,twocolumn,superscriptaddress]{revtex4-2}
\usepackage{amsmath}
\usepackage{graphicx}
\usepackage{subfigure}
\usepackage{xcolor}
\usepackage{amsfonts}
\usepackage{hyperref}
\usepackage{bigstrut,multirow,rotating}
\usepackage{appendix}
\usepackage{svg}

\newcommand\reply[1]{{{{ #1}}}}
\def\cA{$\mathcal{A}$}
\def\cB{$\mathcal{B}$}
\def\cC{$\mathcal{C}$}
\def\cD{$\mathcal{D}$}
\def\cE{$\mathcal{E}$}

\begin{document}

\title{Phase diagrams of $S=\frac 12$ bilayer Models of SU(2) symmetric antiferromagnets }
\author{Fan Zhang}
\affiliation{School of Physics and Astronomy, Beijing Normal University, Beijing 100875, China}
\author{Nisheeta Desai}
\affiliation{Tata Institute for Fundamental Research, Colaba, Mumbai-400 005, India.}
\author{Wenan Guo}
\email{waguo@bnu.edu.cn}
\affiliation{School of Physics and Astronomy, Beijing Normal University, Beijing 100875, China}
\affiliation{Key Laboratory of Multiscale Spin Physics (Ministry of Education), Beijing Normal University, Beijing 100875, China}
\author{Ribhu K. Kaul}
\email{ribhu.kaul@psu.edu}
\affiliation{Department of Physics, The Pennsylvania State University, University Park, PA-16802, USA}
\date{\today}

\begin{abstract}
We study the $T=0$ phase diagrams of models of bilayers of $S=1/2$ square lattices antiferromagnets with SU(2) Heisenberg symmetry that have 2, 4, and 6 spin exchanges.  We study two families of bilayer models with distinct internal symmetries and, hence, different phase diagram topologies. A traditional bilayer model in which the interlayer interaction is Heisenberg so that the two layers can exchange spin (and energy) with each other, making it possible to achieve a simple dimerized valence bond liquid-like state. The resulting phase diagram is rich with N\'eel, valence bond solid and simple dimer phases, and both first-order and continuous transitions, which we demonstrate are consistent with the conventional Landau theory of order parameters. In the second family of models in which the layers can exchange only energy but no spin (reminiscent of the Ashkin-Teller coupling), the simple dimer state cannot occur. The phase diagrams reveal a number of phase transitions that are accessed for the first time. We find that the phase transition between N\'eel and VBS is first order in both the spin-spin and energy-energy coupled models, although they have strikingly distinct finite-size scaling behavior and that the transition from VBS to dimer in the spin-spin coupling model deviates from the expected scenario of an XY model with dangerously irrelevant four-fold anisotropy.     
\end{abstract}

\maketitle

\section{Introduction}\label{sec:ins}

The study of phase diagrams of quantum spin models has led to significant insights into many body physics and quantum field theories. An important issue in such spin models is the study of quantum phase transitions tuned by non-thermal parameters and thus $T=0$ analogs of the familiar thermal phase transition \cite{Sachdev_2011, Sachdev_2008}. A complication that makes the quantum phase transitions in quantum spin models different from thermal phase transitions in classical spin models is the Berry phases of quantum spins, which gives rise to novel topological terms in the $d+1$ dimensional quantum to classical mapping. These terms do not have a classical analog and lead to surprising behavior in quantum spin models. Well-known examples of such phenomena are the distinction between even and odd half-integer spin chains in $1+1$ dimensions~\cite{Haldane_1983, Haldane_1988} and the proposal of deconfined criticality in $2+1$ dimensions~\cite{Senthil_2004, Senthil_2004_2}. While spin chains can be studied using a variety of analytical~\cite{sutherland2004:bm,giamarchi2004:one} and numerical methods~\cite{White_1992}, two and higher-dimensional spin models are still poorly understood because of the lack of reliable methods with which they can be simulated or studied. The proposal of deconfined criticality has spurred extensive studies of quantum phase transitions in higher dimensions, and quantum Monte Carlo (QMC) simulations of sign problem-free Hamiltonians have played an indispensable role here~\cite{Kaul_Melko_Sandvik_2013}.

In this work, we investigate the quantum phase transitions of bipartite SU(2) quantum spin models in a bilayer 
geometry. The bilayer geometry was popularized as one of the first two-dimensional (2D) spin models in which a 
quantum phase transition could be studied unbiasedly using QMC simulation ~\cite{Sandvik_1994, wang_prb2006, Kaul_2012_2}.
Since that early work, there are new models in which the destruction of anti-ferromagnetism (AFM) can be carried out in a single layer, namely using the four-spin $Q$ interaction and its extensions~\cite{Sandvik_2007, Lou_2009}. 
These multi-spin interactions have been instrumental in the studies of AFM to columnar valence bond solid (VBS) 
transitions using QMC simulations. 
The nature of the phase transition in the JQ model in a single layer has been controversial since the early studies~\cite{Sandvik_2007, Melko_2008, Jiang_2008}. 
Although significantly modified scaling can explain numerical results that deviate from simple scaling behavior for continuous transition up to rather large lattices\cite{Shao_2016, Sandvik_2010_2}, recent work has collected compelling evidence of a very weak first-order transition~\cite{demidiofirst,takahashi2024so5, DengPRL}. We do not study the single layer N\'eel-VBS phase transition further in this work, instead we focus on the nature of the related phase transitions in the bilayer. 

\begin{figure}[t]
\centering
\setlength{\abovecaptionskip}{20pt}
\includegraphics[trim=2cm 1cm 1.8cm 0cm, clip,angle=0,width=0.48\textwidth]{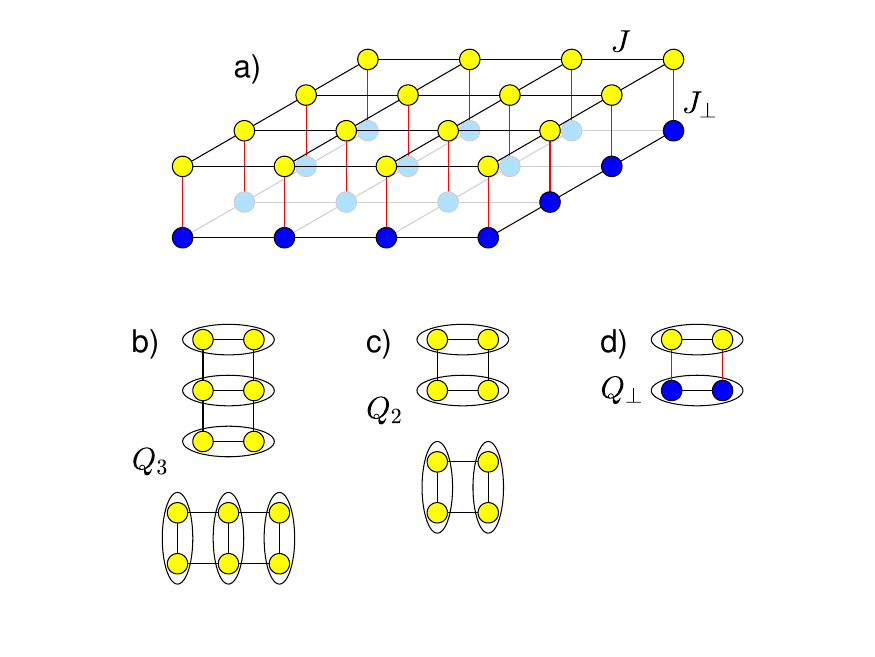}
\caption{The bilayer lattice and its interactions. Yellow spheres represent the spins in the first layer, while blue spheres denote the spins in the second layer. The thin black lines in (a) indicate the intralayer exchange interactions
denoted as $J$, whereas the thin red lines in (a) represent the interlayer S-S exchange interactions 
denoted as $J_\perp$. The ellipses depict singlet projection interactions. (b) depicts the intralayer six-spin interaction, 
denoted as $Q_3$. (c) depicts the intralayer four-spin interaction
denoted as $Q_2$. (d) depicts the interlayer four-spin interaction 
denoted as $Q_\perp$. }
\label{lattice}
\end{figure}

In this work, we combine the bilayer geometry with the six-spin and four-spin $Q$ models to study the generalized 
phase diagram that can occur, focusing on the nature of the associated phase transitions. 
All the models studied here are on square lattice bilayer geometry shown in Fig.~\ref{lattice}(a). Although not illustrated in the figure, it is clear that the lattice sites in the bilayer geometry can be divided into A and B sublattices with all nearest neighbors (both within a layer and on different layers) on opposite sublattices.  All the interactions we 
study are constructed from the $S=\frac 12$ spin singlet projector, 
\begin{equation}\label{Pij}
P_{i,j}=\frac 14 -\vec S_{i}\cdot \vec S_{j},
\end{equation}
with $i$ and $j$ chosen on opposite sublattices. The SU(2) invariant interaction terms can all be expressed simply in terms of $P_{i,j}$. The interactions (illustrated in Fig.~\ref{lattice}) we use are $J$ (intra-layer Heisenberg), $J_\perp$ (inter-layer Heisenberg), $Q_2$ (intra-layer four-spin), $Q_3$ (intra-layer six-spin) and $Q_\perp$ (inter-layer four-spin) interactions,
\reply{whose explicit forms are defined later in Eq.~(\ref{eq:SS3}) and Eq.~(\ref{eq:EE3})}.  
The models we study here can be divided into two classes, S-S and E-E, depending on how the two layers are coupled to each other (spin-spin or energy-energy). The S-S coupling models have only the usual global SU(2) symmetry associated with the total (of both layers) spin conservation. As we shall discuss in the E-E coupling models, we have an SU(2)$\times$SU(2) symmetry, because the total spin of each layer is individually conserved. This difference in symmetry plays an important role in the resulting phase diagrams.

The paper is organized as follows. In Sec.\ref{sec:model}, we introduce our numerical method and physical observables.
In Sec.~\ref{sec:sscplg}, we study the phase diagrams and various phase transitions of the bilayer model with S-S coupling, incorporating additional six-spin interactions. We select four representative cuts to illustrate the characteristics of the phase diagram in detail.  In Sec.~\ref{sec:eecplg}, we present two bilayer models featuring E-E coupling between the layers and investigate the N\'eel-VBS transition within this framework.  In sec.~\ref{sec:sum}, we provide a comprehensive summary and offer insights into potential future research directions.

\section{Methods and Observables}\label{sec:model}

In this paper, we employ the stochastic series expansion (SSE) QMC method\cite{Sandvik1991-QMC, Sandvik1999-SSE}. Periodic boundary conditions are applied. 
Unless otherwise noted, we have set the inverse temperature $\beta=L$ with $L$ being the linear
size of the system.
\reply{We focus on quantum critical points characterized by a dynamic critical exponent $z=1$, implying that the energy gap vanishes as $\Delta \sim L^{-z} = L^{-1}$. In the finite-size scaling analysis, the inverse temperature $\beta$ must scale with the system size $L$ as $\beta \propto L^z$ to ensure that the system remains in the ground-state regime as the thermodynamic limit is approached. Therefore, we fix the ratio $\beta/L = 1$ throughout our simulations to maintain a constant aspect ratio in the $(2+1)$-dimensional space-time scaling limit.}
Typically $10^8$ MC samples are taken for each set of parameters.

To characterize the phases that appear in our models, we use a number of different observables. The magnetically ordered phase with the O(3) spin rotational symmetry broken can be characterized by the $z$ component of the order parameter  of N\'eel state, which is defined as:
\begin{equation}\label{eq:ms}
{m^z_s}(\ell)=  \frac 1N \sum_{{\vec r \in \ell}} S_{\ell \vec r}^z e^{-i\vec k\cdot \vec r}
\end{equation}
with $\vec k=(\pi,\pi)$ is the wave vector corresponding to the N\'eel phase. $\ell=1,2$ is the layer index. $m^z_s$ 
is diagonal in the $S^z$ basis, making it straightforward to sample within the SSE representation of QMC. For convenience, we will omit the superscript $z$. $\langle m_s^2 \rangle$ is finite in a N\'eel ordered phase and zero in a N\'eel disordered phase in the thermodynamic limit. The Binder cumulant of the N\'eel order parameter is defined as:
\begin{equation}\label{eq:Um}
    U_{m}(\ell)=\frac 52 \left(1-\frac 13 \frac{\langle m_s^4(\ell)\rangle }{\langle m_s^2(\ell)\rangle ^2}\right ),
\end{equation}
where $\ell$ is also the layer index. $U_m$ is a dimensionless quantity that proves to be very useful for studying phase transitions, particularly in locating the phase transition point and distinguishing between continuous and first-order phase transitions. As $L \to \infty$, $U_m$ approaches 1 in the ordered phase and 0 in the disordered phase. 
The crossing value of the tuning parameter for different system sizes will converge to the phase transition point as $L \to \infty$.
In the case of a first-order phase transition, the squared order parameter follows a bimodal distribution, therefore, $U_m$ demonstrates a negative divergent peak, as the system size $L$ 
approaches infinity near the phase transition point\cite{Binder-firstorder}.

The order parameter of the VBS state, which breaks the $Z_4$ symmetry of the lattice translations, can be characterized by a two-component vector $\vec \phi=(\phi_x,\phi_y)$
with
\begin{equation}\label{eq:phi}
\begin{split}
&\phi_x(\ell) =\frac 1N \sum_{\vec r \in \ell} S^z_{\ell \vec r}S^z_{\ell (\vec r +\hat x)} e^{-i\vec k\cdot \vec r} , \\
&\phi_y(\ell)=\frac 1N \sum_{\vec r \in \ell}S^z_{\ell \vec r}S^z_{\ell (\vec r +\hat y)} e^{-i\vec k\cdot \vec r},
\end{split}
\end{equation} 
where $\hat x$ and $\hat y$ are unit vectors along the $x$ and $y$ direction, respectively. The wave vectors are 
$\vec k=(\pi,0)$ for $\phi_x$ and $(0,\pi)$ for $\phi_y$. $\langle \phi^2 \rangle$ is a finite value in the VBS 
ordered phase and 0 in the VBS disordered phase in the thermodynamic limit. Based on the VBS order parameter, we 
also define a corresponding Binder cumulant of the VBS state:
\begin{equation}\label{eq:Uphi}
U_{\phi}(\ell)=2\left(1-\frac 12 \frac{\langle \phi^4(\ell)\rangle }{\langle \phi^2(\ell)\rangle ^2}\right).
\end{equation}
$U_\phi$ approaches 1 in the VBS-ordered phase and 0 in the VBS-disordered phase. The behavior of $U_\phi$ in 
detecting VBS order is similar to that of $U_m$ when it comes to identifying N\'eel order.

Due to the symmetry between the two layers, the behavior of the order parameters and Binder cumulants is 
identical for both layers. Therefore, in the following, we present the results of quantities for one layer as representative of both, unless specifically noted otherwise.

Finally, the spin stiffness $\rho_s$ is defined by 
\begin{equation}
\begin{split}
 \rho_s=\frac 1N \frac{\partial^2F(\varphi)}{\partial \varphi^2},
\end{split}
\end{equation}
where $F$ is the free energy and $\varphi$ is the twisted angle. The spin stiffness, $\rho_s$, is also a valuable indicator for detecting magnetic order. It is a finite value in the magnetically ordered or quasi-long-range ordered phase and approaches zero in the disordered phase. At the critical point, $\rho_s$ scales as\cite{Fisher_boson}:
\begin{equation}
\rho_s \sim L^{2-d-z},
\end{equation}
where $d$ is the dimensionality of space and $z=1$ is the dynamic critical exponent. Therefore, $L\rho_s$ is also a dimensionless quantity and will behave similarly to $U_m$.
The spin stiffess can be sampled in the SSE QMC by the relation:
\begin{equation}
\rho_s=\frac {3}{4\beta} \left( \langle W_x^2\rangle + \langle W_y^2 \rangle \right),
\end{equation}
where $W_x$ and $W_y$ are winding numbers of spin transporting in the $x$ and $y$ directions,
respectively.

\section{S-S Coupling}
\label{sec:sscplg}

In this section, we will address the phase diagram and phase transitions in the bilayer $S=\frac 12$ anti-ferromagnets in which the interlayer exchange, being of the spin-spin type, allows spin and energy to be exchanged between the layers. 

\subsection{Model and Phase Diagram}

\begin{figure}[h]
\centering
\setlength{\abovecaptionskip}{20pt}
\includegraphics[angle=0,width=0.48\textwidth]{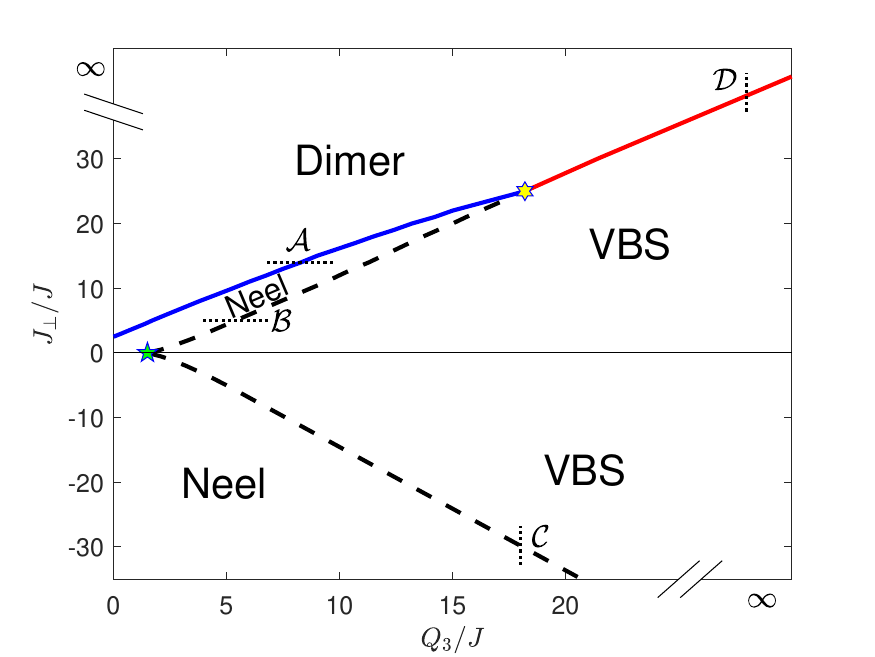}
\caption{Phase diagram of bilayer S-S coupling model, Eq.~\ref{eq:SS3}. We have constructed the phase boundaries by extensive simulations of the model on moderate system sizes. 
  In addition to probing the nature of the phase transitions, we have studied four cuts $\mathcal{A}$ ($J_\perp/J=14$), $\mathcal{B}$ ($J_\perp/J=5$), $\mathcal{C}$ ($J=0, J_\perp<0$), $\mathcal{D}$ ($J=0, J_\perp>0$) in detail on large lattices. Briefly, we find that the phase transition on the $\mathcal{A}$ cut belongs to 3D O(3) universality, whereas $\mathcal{B}$ and $\mathcal{C}$ cuts are first-order phase transitions.  All of these are consistent with the expectations of a straightforward Landau theory. Most interesting is $\mathcal{D}$ cut, where we observe behavior consistent with a continuous phase transition. One natural scenario for a continuous transition is the three-dimensional XY transition with a dangerously irrelevant four fold magnetic field anistropy~(for recent work see e.g. \cite{carmona6loop,Shao_2020}).  However, from our numerical simulations we find in our model a persistent $Z_4$ anisotropy that contradicts the expected emergent O(2) symmetry of the dangerously irrelevant scenario. We are unable to offer a consistent theoretical scenario for this numerical observation. In our phase diagrams, we represent first-order phase transitions with dashed lines and continuous phase transitions with solid lines. The blue line indicates the O(3) phase transition, whereas the universality class of the red line is not clear yet. 
  Additionally, two stars are marked: the blue star denotes the single-layer N\'eel-VBS transition, 
  and the yellow star represents a multi-critical point where the three phases meet.}
\label{jq3jpd}
\end{figure}

In this subsection, we introduce the model Hamiltonian for the bilayer with spin-spin interlayer coupling and give an overview of its phase diagram. We begin our study with the traditional bilayer  Heisenberg model with the in-plane interaction $J$ and the rung interaction $J_\perp$ exchange ~\cite{Sandvik_1994}. When $J>0, J_\perp>0$, the interactions of both inter and intra layers are antiferromagnetic. This system hosts a N\'eel state for $J
\gg J_\perp$ and a simple dimer state when $J \ll J_\perp$. The dimer state is smoothly connected to a trivial product state. Hence, the quantum phase transition between N\'eel and simple dimer is expected to be the conventional 3D O(3) universality, a fact that 
has been tested with high precision through QMC simulations~\cite{wang_prb2006}. A 
ferromagnetic $J_\perp<0$ only strengthens the N\'eel state (this is shown on the $y$-axis in 
Fig.~\ref{jq3jpd}). It is interesting to ask how this phase diagram accommodates a VBS state, which is induced by large $Q_3$.

To this end, we introduce the $J$-$Q_3$ spin-spin (S-S) coupling Hamiltonian, which is written as,
\begin{equation}
\begin{split}
\label{eq:SS3}
H_{SS3}=& -J\sum_{\ell,\langle ij\rangle } P_{\ell i,\ell j} -J_\perp \sum_{ i  } P_{1i,2i} \\
&- Q_3 \sum_{\ell, \langle ijklmn\rangle} P_{\ell i,\ell j}P_{\ell k,\ell l}P_{\ell m,\ell n},
\end{split}
\end{equation}
where $\langle ij\rangle$ denotes nearest neighbors on a square periodic lattice, as illustrated in Fig.~\ref{lattice}~(a), which consists of $N = L^2$ sites per layer, with $\ell = 1, 2$ representing the layer index. The notation $\langle ijklmn\rangle$ refers to a $2 \times 3$ and $3 \times 2$ $Q_3$ plaquette in the $J$-$Q_3$ model, depicted in Fig.~\ref{lattice}~(b). The summation over $i$ encompasses the $L^2$ sites of the square lattice shown in Fig.~\ref{lattice}.

\begin{figure}[t]
\centering
\setlength{\abovecaptionskip}{20pt}
\includegraphics[trim=1.2cm 3cm 1cm 1cm,clip,angle=0,width=0.48\textwidth]{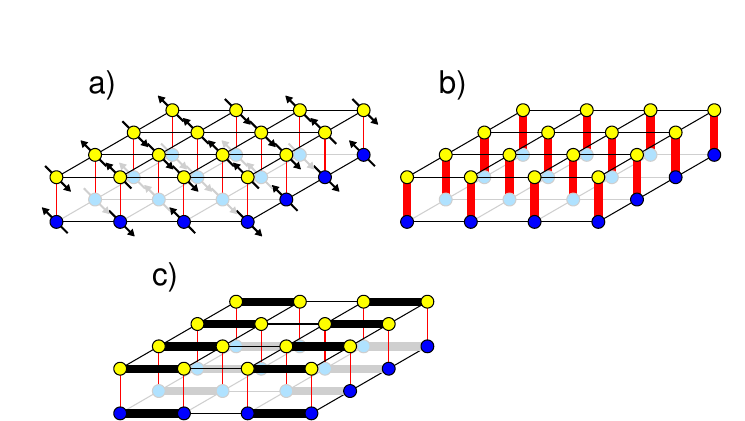}
\caption{The cartoons of phases of this model. a) N\'eel phase. b) Dimer phase. 
c) VBS phase. Arrows in (a) represent the direction of spins. However, it is important to note that the N\'eel state in a quantum model is not identical to the antiferromagnetic phase in a classical model. This difference arises because, in the quantum model, spins are not arranged regularly due to quantum fluctuations in the ground state. Thick bonds in b) and c) represent singlet states, which are $\frac 1{\sqrt{2}} (| \uparrow \downarrow \rangle -| \downarrow \uparrow \rangle)$. The VBS state breaks the $Z_4$ translational symmetry of the lattice. It has four distinct ground states. Figure (c) illustrates one of these ground states. }
\label{phase}
\end{figure}

Before delving into a detailed exploration of the phase transitions in the spin-$\frac{1}{2}$ S-S bilayer $J$-$Q_3$ model, we first present the phase diagram of this model, as illustrated in Fig.~\ref{jq3jpd}. This diagram is derived from the analysis that will be discussed in the following sections. The model exhibits three distinct phases: the simple dimer phase (shown in Fig.~\ref{phase}~(b)), the N\'eel state (depicted in Fig.~\ref{phase}~(a)), and the valence bond solid (VBS) state (illustrated in Fig.~\ref{phase}~(c)).  The simple dimer state preserves all symmetries, and the VBS breaks some lattice symmetries; both of these states maintain the spin rotational symmetry, which is broken in the N\'eel phase. We now turn to the study of the phase transitions between these phases. We present numerical data along four representative cuts, which are labeled $\mathcal{A}$, $\mathcal{B}$, $\mathcal{C}$, and $\mathcal{D}$ in Fig.~\ref{jq3jpd}.

\subsection{Phase Transitions on the  \cA, \cB, \cC, \cD~cuts}

We now study the nature of the phase transitions along the four cuts \cA, \cB, \cC~ and \cD, in the
S-S bilayer model, as indicated in Fig.~\ref{jq3jpd}.

\subsubsection{Cut \cA}

\begin{figure}[t]
\centering
\setlength{\abovecaptionskip}{20pt}
\includegraphics[angle=0,width=0.48\textwidth]{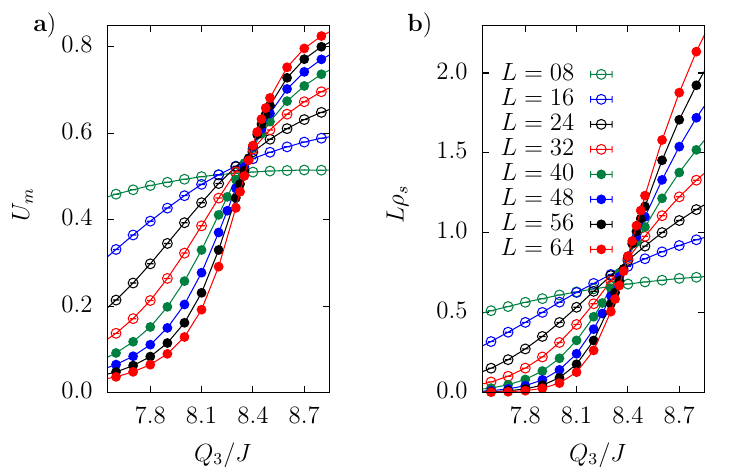}
\caption{The finite-size behavior of $U_m$ and $\rho_sL$ at cut \cA. (a) shows $U_m$ varies with $g_\mathcal{A} = Q_3/J$ near the phase transition point for different system sizes $L$. (b) illustrates how $L\rho_s$ changes with $g_\mathcal{A}$ near the phase transition for various system sizes $L$. Both tend to 0 in the disordered phase and approach finite values in the N'eel ordered phase, with their crossings converging to the phase transition point as the system size $L$ increases. }
\label{A-binder}
\end{figure}

\begin{figure}[h]
\centering
\setlength{\abovecaptionskip}{20pt}
\includegraphics[angle=0,width=0.48\textwidth]{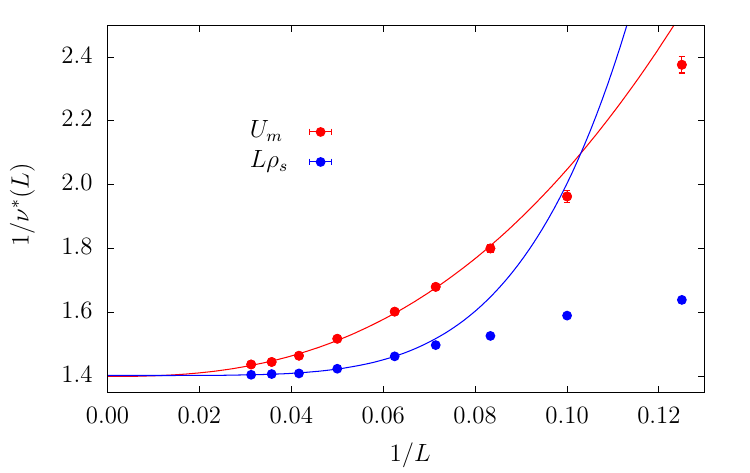}
\caption{The finite size scaling study of $1/\nu$ by the crossing points analysis of $(L,2L)$ on cut \cA~of Fig.~\ref{jq3jpd} where $J_\perp/J=14$. The red color refers to the crossing points from $U_{m}$. Blue colors refer to the results from spin stiffness. Filled circles refer to the QMC data, and the solid lines are fits using Eq. (\ref{eq:nuc}). 
We find 
$1/\nu=1.40(1)$ 
from $U_m$ and $1.403(2)$
from $L\rho_s$. The fitting windows is $L=12 \sim 32$ for $U_m$, $L=16 \sim 32$ for $L\rho_s$.}.
\label{A-1nu}
\end{figure}

Along the line $J_\perp/J=14$, the phase transition on cut \cA~ is expected to be in the universality class of the 3D O(3) sigma model. To verify this behavior, here we extract the critical exponent $\nu$ from our data and compare it with the known value for the O(3) universality class.  
 This is done by applying the general finite-size scaling\cite{Nightingale, Fisher_Baber} of a dimensionless quantity $A$ 
 near the critical point.

Let $g$ be the tuning parameter that drives the phase transition, with $g_c$ representing the critical point. For a dimensionless quantity $A$, the crossing point $g^*(L)$ for finite sizes $(L, rL)$ will converge to $g_c$ according to a power law:
\begin{equation}
    g^*(L)-g_c = a L^{-1/\nu+\omega} 
    \label{gc}
\end{equation}
where $\nu$ is the correlation length exponent, $\omega> 0$ is the leading irrelevant exponent, and $a$ an unknown coefficient.
At the crossing point $g^*(L)$, an finite size exponent estimate $\nu^*(L)$ for correlation length exponent $\nu$
can be defined by the ratio
of the slopes of the two curves
\begin{equation}\label{eq:nu}
{1}/{\nu^*(L)}=\frac{1}{\ln(r)} \ln\left( \frac{dA(g^*,rL)/dg} {dA(g^*,L)/dg}\right),
\end{equation}
which also converges to $1/\nu$ by power law:
\begin{equation}
    {1}/{\nu^*(L)}-{1}/{\nu} = b L^{-\omega},
    \label{eq:nuc}
\end{equation}
with $b$ an unknown coefficient. 
For details of the crossing analysis, see, e.g., the Supplemental Material of Ref. \cite{Shao_2016}.


With $J_\perp/J=14$, the Binder cumulant $U_{m}$ of N\'eel order parameter and 
the scaled spin stiffness $L \rho_s$ as functions of $g_\mathcal{A}=Q_3/J$ are calculated for different system sizes.  The results are shown in Fig.\ref{A-binder}. The two quantities are dimensionless. The crossing points $g^*_\mathcal{A}(L)$ of $U_{m}$ for system sizes $L$ and $2L$  are calculated and analyzed using Eq. (\ref{gc}). We obtain $g_{\mathcal{A}c}=8.1(1)$.  
Similar analysis to the $L\rho_s$ data leads to consistent results.

The exponent estimate $\nu^*(L)$ at the crossings of $U_m$ is also determined and analyzed using Eq. (\ref{eq:nuc}). We obtain $1/\nu = 1.40(1)$. The crossing analysis for $L\rho_s$ yields $1/\nu = 1.403(2)$. These consistent results align with the 3D O(3) value \cite{nu_ref}. The details of the analysis are presented in Fig. \ref{A-1nu}.

\subsubsection{Cut \cB}

We now turn to a study of the phase transition between N\'eel and VBS phases in the bilayer geometry. Because of the interlayer spin-spin coupling, it has been argued that the spin Berry phases cancel between the two layers, making a conventional Landau theory applicable~\cite{Sachdev_2011,Senthil_2004_2}. In a conventional Landau theory, a generic direct transition between two ordered phases with distinct order 
parameters must be first order. We thus expect the N\'eel-VBS transition in the bilayer geometry to be first order as found in previous studies of SU($N$) magnets~\cite{Kaul_2012_2}. We verify that this is indeed the case by looking at the histogram of the VBS order parameter $\vec \phi$.

\begin{figure}[t]
\centering
\setlength{\abovecaptionskip}{20pt}
\includegraphics[angle=0,width=0.48\textwidth]{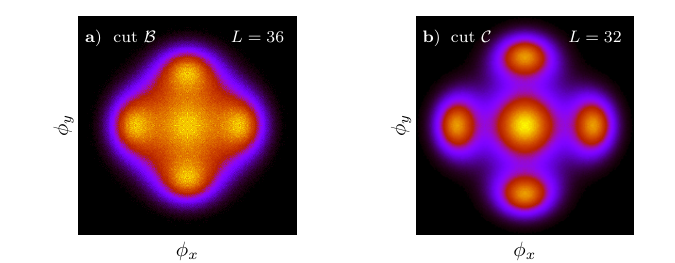}
\caption{
	Histogram of the VBS order parameter $\vec{\phi}$ of one layer at cut \cB~ (a) and cut \cC~(b). The brighter color indicates a higher probability of occurrence, with the $x$ and $y$ axes representing $\phi_x$ and $\phi_y$ ranging from $-0.1 \sim 0.1$. 
	(a) histogram for system size $L=36$ at $Q_3/J= 5.462$ with $J_\perp/J=5$. 
	(b) histogram for system size $L=32$ at $J_\perp/Q_3= -2.224$ with $J=0$. 
	The diagrams show the coexistence of N\'eel state (central peak at origin) and VBS state (four symmetric peaks) at cut \cB~ and \cC, respectively, indicating the transitions are first-order, as expected by the conventional Landau theory.}
\label{BC-his}
\end{figure}

Along the line $J_\perp/J=5$, the N\'eel-VBS transition is located near $Q_3/J \sim 5.462$, called cut \cB.
This is evident from the histogram of the VBS order parameter $\vec \phi$, illustrated in Fig.\ref{BC-his}~a). 
The histogram shows the probability distribution of $\vec \phi$:
The brighter color indicates a higher probability of occurrence, with the $x$ and $y$ axes representing $\phi_x$ and $\phi_y$, respectively. Bright spots appear at the center and finite values on the $x$ and $y$ axes, indicating the coexistence of the VBS disordered phase and the $Z_4$ ordered phase near the phase transition point, a characteristic of a first-order phase transition. 

\begin{figure}[t]
\centering
\setlength{\abovecaptionskip}{20pt}
\includegraphics[angle=0,width=0.48\textwidth]{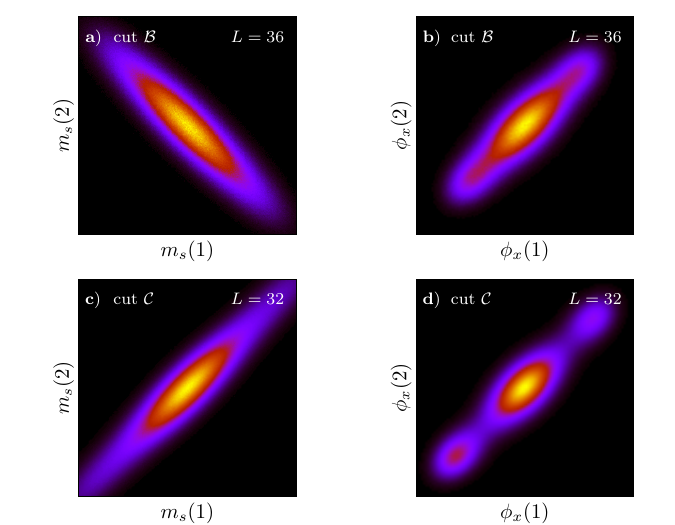}
\caption{ Histogram of N\'eel order parameter $(m_s^z(1), m_s^z(2))$ and VBS order parameters $(\phi_x(1), \phi_x(2))$  at cut \cB~ and \cC. The $m_s^z$ range range from $-0.2 \sim 0.2$ and $\phi_x$ range from $ -0.1 \sim 0.1$. (a) and (b) are histograms for system size $L=36$ at cut \cB. 
(c) and (d) are histograms for system size $L=32$ at cut \cC. 
All of the histograms are diagonal, which means that the orders of the two layers are locked together.}
\label{BC-his2}
\end{figure}

As previously mentioned, due to the coupling between the two layers, the expectation values of the order parameters in both layers are consistent; however, the microscopic states of the two layers may not be completely identical, which we explore through histograms. 
Figure \ref{BC-his2} (a) shows the histogram of vector $(m_s^z(1), m_s^z(2))$ formed by N\'eel order parameter of layer 1 and layer 2. 
the horizontal and vertical axes represent $m_s^z(1)$ and $m_s^z(2)$, respectively. The bright stripe along the 
diagonals in the second and fourth quadrants indicates that the two order parameters are opposite-directed, as 
expected from the AF S-S coupling between the two layers. 
Figure \ref{BC-his2} (b) shows the histogram of a vector $(\phi_x(1), \phi_x(2))$ formed by the $x$-components of
the VBS order parameter in layer 1 and layer 2. 
The horizontal and vertical axes represent $\phi_x(1)$ and $\phi_x(2)$, respectively. The bright stripe appearing 
along the diagonal in the first and third quadrants indicates that the VBS states of the two layers are in phase.

\subsubsection{Cut \cC}

We now turn to the study of the N\'eel-VBS transition on the ferromagnetic side of the spin-spin coupling 
$J_\perp <0$.
Without $Q_3$ interactions, the FM interlayer exchange $J_\perp$ strengthens the N\'eel order. It is quite natural 
that including strong multispin $Q_3$ interactions drives the system into a VBS phase. Such a transition continuous
to strong $J_\perp$, where 
the spins at site $i$ of two layers, $S_{1i}$ and $S_{2i}$, tend to form  $S=1$ spin triplets.  
Therefore, the N\'eel-VBS transition line is expected to be described by an effective $S=1$ model, making the
conventional Landau theory applicable. The transition is then expected to be first-order. 

We here show numerical evidence by analyzing the histogram of the VBS order parameter $\vec \phi$ at cut \cC 
that the phase transition is first order, in agreement with the QMC simulations of the N\'eel-VBS transition of 
a $S=1$ model\cite{Wildeboer_2020}.
The histogram of the VBS order parameter $\vec \phi$ at cut \cC~ is presented in Fig.~\ref{BC-his}~(b). 
The coexistence of the VBS disordered state, i.e., the N\'eel state, the $Z_4$ ordered VBS state is evident,
confirming the discontinuous nature of the transition. 

Additionally, we sampled histograms of the
N\'eel and VBS order parameters across different layers, illustrated in Fig.~\ref{BC-his2}~(c) and (d). The 
results demonstrate that both the N\'eel order and the VBS order in the two layers are locked, exhibiting 
precisely the same orientations for both order parameters.

\subsubsection{Cut \cD}
\begin{figure}[t]
\centering
\setlength{\abovecaptionskip}{20pt}
\includegraphics[angle=0,width=0.48\textwidth]{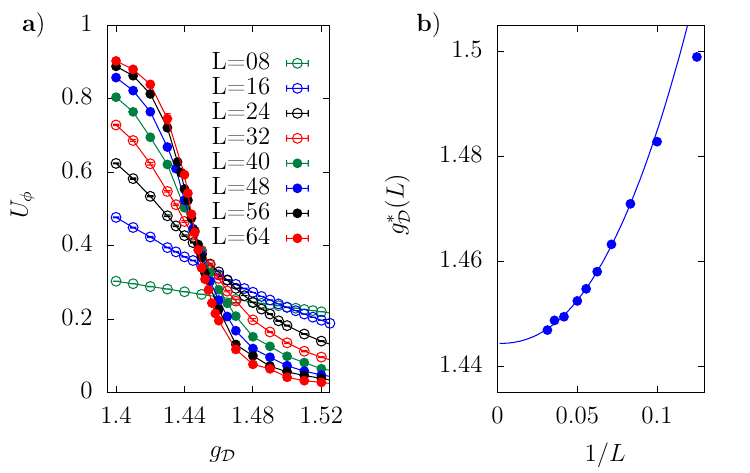}
\caption{ The behavior of $U_\phi$ and fit of the phase transition point at cut \cD.
(a) $U_\phi$ vs. $g_\mathcal{D}=J_\perp/Q_3$ for different system sizes. 
The normal behavior of the Binder cumulant implies a continuous phase transition. 
(b) shows crossing points $g_D^*(L)$ of $U_\phi$ curves for system size pair $(L, 2L)$ as function of $1/L$. 
Fitting Eq. (\ref{gc}) to $g_\mathcal{D}^*(L)$ with  $L=12\sim 36$ leads to  $g_{\mathcal{D}c}=1.4443(6)$. }
\label{B-binder}
\end{figure}

Finally, we turn to study the phase transition along the cut $\mathcal{D}$ depicted in Fig.~\ref{jq3jpd}.  
The cut sits at the limit of the in-plane Heisenberg coupling $J=0$. There is no room for the N\'eel phase. The 
\reply{competition}
between $J_\perp$ and $Q_3$ leads to a phase transition between the Dimer phase and the VBS phase, which
can be investigated using the finite-size behavior of the VBS Binder cumulant $U_\phi$. 

Figure~\ref{B-binder} (a) illustrates finite-size behavior of $U_\phi$ as a function of 
$g_\mathcal{D}=J_\perp/Q_3$ near the transition at cut \cD, which shows typical behavior of a continuous phase
transition. Crossing point analysis is applied to extract the critical point to $g_{\mathcal{D}c}=1.4443(6)$, as 
shown in Fig.~\ref{B-binder}~(b).

This transition marks the change from a VBS phase, which breaks the lattice translation symmetry, to a trivial dimer phase 
in which all symmetries are restored. 
From a general perspective of the Landau theory, this transition is anticipated to show the same critical behavior as the classical 3D 4-state clock model. 
The characteristics of the phase transition in the 3D 4-state clock model are well understood and are expected to exhibit 3D O(2) or U(1) universality, albeit with a “dangerously irrelevant” $Z_4$ perturbation\cite{cft-cardy1996scaling}. 
Consequently, just near the phase transition, we expect to observe interesting crossover phenomena. 
Namely, the histograms of $\vec{\phi}(l)$ for a single layer on the VBS side but close to the critical point should
exhibit an emergent U(1) symmetry, which gradually transitions into a $Z_4$ symmetric 
distribution as the lattice size increases, as demonstrated in the clock model \cite{Lou_clock, Shao_2020}.

\begin{figure}[t]
\centering
\setlength{\abovecaptionskip}{20pt}
\includegraphics[angle=0,width=0.48\textwidth]{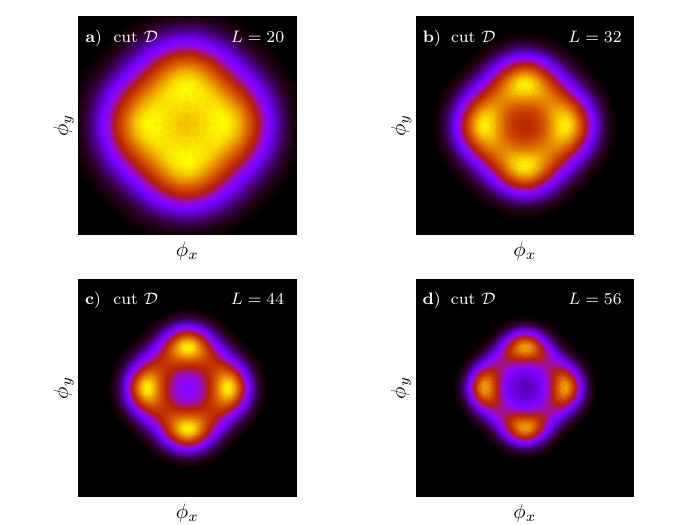}
\caption{
The histogram of VBS order parameter $(\phi_x,\phi_y)$ near cut \cD, which is $J=0, J_\perp/Q=1.430$, close to the critical point but just in the VBS side, for different system sizes: a) $L=20$, b) $L=32$, c) $L=44$, d) $L=56$.
All histograms exhibit $Z_4$ anisotropy.
}
\label{D-his}
\end{figure}

In Figure \ref{D-his} the histogram of $\vec \phi$ in the VBS side close to the transition point is shown. 
Interestingly, the histograms for different system sizes all exhibit $Z_4$ anisotropy. We do not observe the 
expected emergent U(1) symmetry.


\begin{figure}[t]
\centering
\setlength{\abovecaptionskip}{20pt}
\includegraphics[angle=0,width=0.48\textwidth]{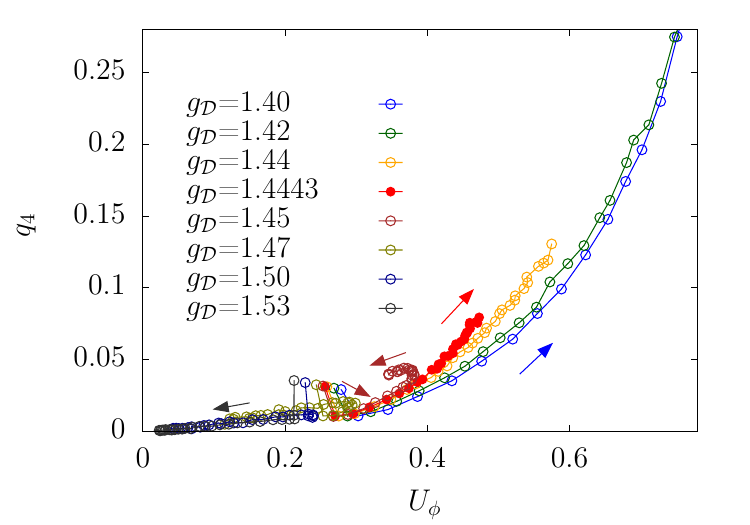}
\caption{The numerical renormalization flow diagram near cut \cD. In this diagram, the x-axis represents the Binder cumulant $U_{\phi}$, while the y-axis corresponds to $q_4$ defined by Eq.(\ref{eq:q4}). A group of markers connected by a solid line indicates the values of $(U_{\phi}, q_4)$ for different system sizes at the same coupling strength $g_\mathcal{D} = J_\perp / Q_3$. The direction of the arrows points towards the increasing system size, and the solid red circles represent the data points at the phase transition point $g_{\mathcal{D}c} = 1.4443$. The sizes used in the diagram range from $L = 6 \sim 64$.}
\label{D-rg}
\end{figure}

 It is worth noting that the $Z_4$ symmetry breaking associated with the VBS order is not entirely the 
 same as the $Z_4$ symmetry breaking in the clock model. The former is related to the symmetry of the lattice, 
 while the latter pertains solely to the symmetry of the spins. To verify whether the $Z_4$ symmetry breaking 
 of the VBS order is relevant at the critical point, we define an angular order parameter $q_4$ to measure the 
 $Z_4$ anisotropy of the system:
\begin{equation}\label{eq:q4}
q_4=\langle \cos (4 \Theta ) \rangle
\end{equation}
with the angle $\Theta=\arctan \frac{\phi_y}{\phi_x}$ defined for each configuration. This quantity becomes 
nonzero if the system is $Z_4$ anisotropy. 

To elucidate the difference between the $Z_4$ anisotropy in the clock model and the current model, we employ the 
numerical renormalization flow diagram analysis, as depicted in 
Fig.~\ref{D-rg}. The $x$ axis represents the Binder cumulant $U_{\phi}$, while the $y$ axis 
corresponds to the angular order parameter $q_4$. The same group of markers connected by a solid line indicates 
the values of $(U_{\mathrm{VBS}}, q_4)$ for different system sizes at the same coupling strength. The direction of 
the arrows points towards increasing system sizes, and the solid red circles denote data at the critical point $g_{\mathcal{D}c} = 1.4443$.
In the VBS phase, with $g_\mathcal{D} <  g_{\mathcal{D}c}$, the flow moves towards the VBS ordered fixed point with $U_{\phi} = 1$ and $q_4 = 1$, consistent with the characteristics of the VBS phase that breaks the $Z_4$ lattice translation symmetry. In the dimer phase, with $ g_{\mathcal{D}} >  g_{\mathcal{D}c} $, the flow heads towards the disordered fixed point with $U_\phi = 0$ and $q_4 = 0$.
 At $g_{\mathcal{D}c}$, the flow approaches a fixed point with both $q_4$ and $U_{\phi}$ finite, differing from 
 the critical flow  of the 3D 4-state clock model (as referenced in FIG.S3 of \cite{Shao_2020}), which  flows to 
 $q_4 =0$ and $U_{\phi}$ remains finite.  
 
 Such a critical flow suggests the $Z_4$ anisotropy is relevant at the VBS-dimer critical point, leading to
 the conclusion that the VBS-dimer transition at cut \cD ~belongs to a universality different from 3D U(1). 
However, this conjecture is not supported by any field theory currently available, 
along with the limitations of numerical methods regarding system sizes, we leave the exploration of the critical 
behavior of this transition for future work. 


\section{E-E Coupling}
\label{sec:eecplg}

\subsection{Model and Phase Diagram}
\label{sec:eempd}

We now turn to the bilayer $S=1/2$ anti-ferromagnets with a different kind of ``energy-energy" (E-E) coupling between the anti-ferromagnetic layers, in which the layers can exchange energy but not spin. As a result, the SU(2) symmetry of each layer is individually preserved -- an interesting consequence is the simple dimer state does not appear. We still have a phase transition between N\'eel and VBS, which we explore. For the individual layers, we have considered both $J$-$Q_3$ and $J$-$Q_2$ models. Since we find that the phase diagrams and phase transitions are very similar, here we only present the $J$-$Q_3$. The Hamiltonian of this bilayer $J$-$Q_3$ E-E coupling model is,
\begin{equation}\label{eq:EE3}
\begin{split}
H_{EE3}=& -J\sum_{\ell,\langle ij\rangle } P_{\ell i,\ell j} -Q_\perp \sum_{\langle ij\rangle } P_{1i,1j}P_{2i,2j} \\
& - Q_3 \sum_{\ell,\langle ijklmn\rangle} P_{\ell i,\ell j}P_{\ell k,\ell l}P_{\ell m,\ell n}.\\
\end{split}
\end{equation}
The E-E coupling is the inter-layer four-spin interaction $Q_\perp$, as illustrated in Fig. \ref{lattice}. Clearly in this form of coupling the individual SU(2) symmetries of each layer is preserved (the layers can still exchange energy) so the physics of this model of coupling is distinct from the S-S coupling. It is expected that under such coupling the N\'eel and VBS phases will be preserved: We note here that we expect from symmetry, that the N\'eel phase is special in the sense that the order parameters in each layer can rotate independently, in contrast we expect the VBS order parameters in each layer to lock even with the introduction of a small E-E coupling -- we will demonstrate both these expectation through our numerical simulations. The remaining question is about the nature of the transition which we will study in detail in the following subsection.   

\begin{figure}[t]
\centering
\setlength{\abovecaptionskip}{20pt}
\includegraphics[angle=0,width=0.48\textwidth]{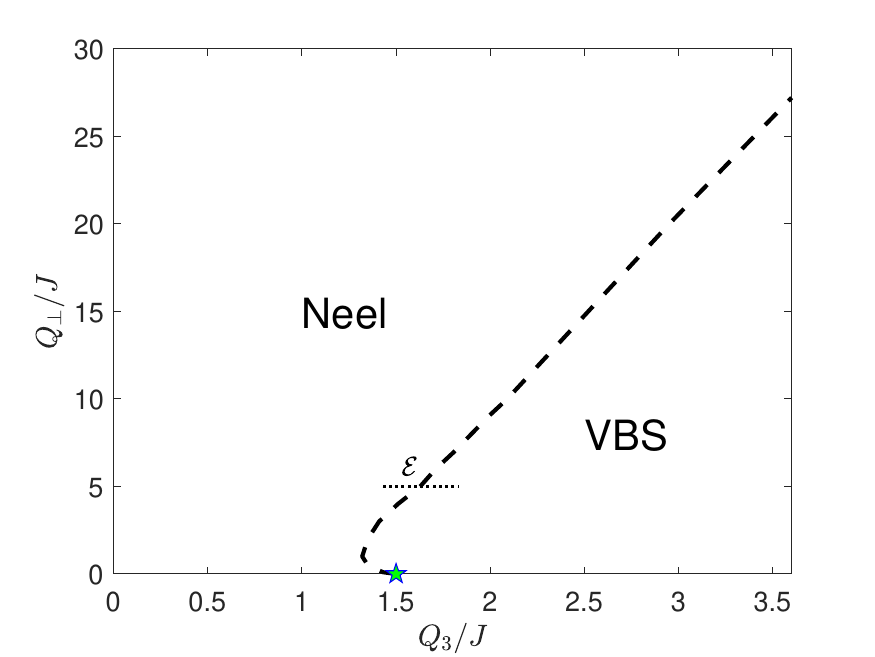}
\caption{Phase diagram of $J$-$Q_3$ bilayer E-E coupling model. There are two phases: the N\'eel phase and the VBS phase. The dotted line shows that the phase transition between them is likely to be first order. The blue star labels the transition point of the 2D $J$-$Q_3$ model. The cut \cE~ along $Q_\perp/J=5$ is well studied below.}
\label{jq3qpd}
\end{figure}

\begin{figure}[htbp]
\centering
\setlength{\abovecaptionskip}{20pt}
\includegraphics[angle=0,width=0.48\textwidth]{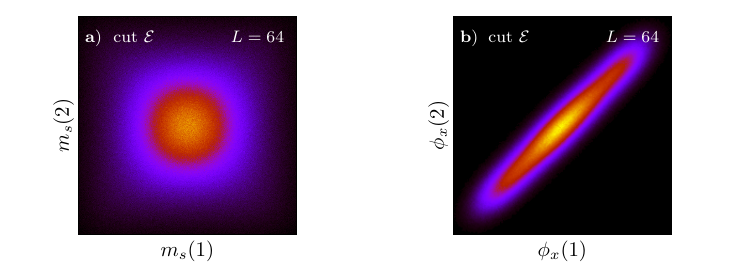}
\caption{Histogram of N\'eel order parameter $(m_s(1), m_s(2))$ (a) and VBS order parameters $(\phi_x(1), \phi_x(2))$ (b) at cut $\mathcal{E}$ for system size $L=64$. The $m_s$ range range from $-0.15 \sim 0.15$ and $\phi_x$ range from $ -0.05 \sim 0.05$. In (a), a round peak is centered at the origin while, in (b), we see a diagonal peak, which means that the N\'eel order of the two layers is independent while the VBS order of the two layers is locked together.}
\label{E-his2}
\end{figure}

The phase diagram of this $J$-$Q_3$ bilayer E-E coupling model is shown in Fig.\ref{jq3qpd}. As expected there are two phases: VBS and N\'eel. We numerically study how the order parameters of the two layers couple through histograms, both in the N\'eel and VBS phases. We show that in both the cases the order parameters couple in precisely the way outlined above.
Figure \ref{E-his2} (a) shows the histogram of vector $(m_s(1), m_s(2))$ formed by N\'eel order parameter of layer 1 and layer 2 at cut $\mathcal{E}$. The horizontal and vertical axes represent $m_s(1)$ and $m_s(2)$, respectively. The bright round peak centered at the origin indicates that the two order parameters are independent, demonstrating that the E-E couplings do not lock the N\'eel order parameters in the two layers.
Figure \ref{E-his2} (b) shows the histogram of $(\phi_x(1), \phi_x(2))$ formed by the $x$-components of the VBS order parameter in layer 1 and layer 2. The horizontal and vertical axes represent $\phi_x(1)$ and $\phi_x(2)$, respectively. The bright stripe appearing along the diagonal in the first and third quadrants indicates that the VBS states of the two layers are locked in phase.

\subsection{Phase Transition on cut \cE}

\begin{figure}[hbtp]
\centering
\setlength{\abovecaptionskip}{20pt}
\includegraphics[angle=0,width=0.48\textwidth]{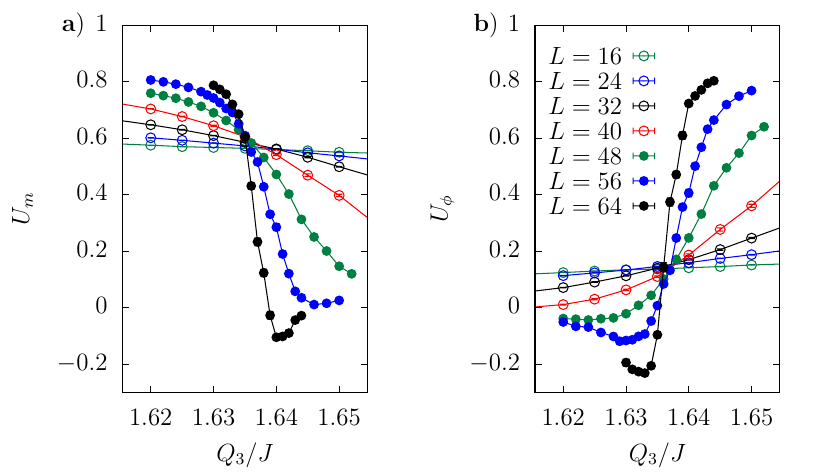}
\caption{The behavior of Binder cumulants (a) $U_m$ and (b) $U_\phi$ v.s. $g_\mathcal{E}=Q_3/J$ at cut \cE~ along $Q_\perp/J=5$.  Both of these two Binder cumulants exhibit negative peaks near the phase transition point when the system sizes are large enough, suggesting the phase transition could be first order.}
\label{E-binder}
\end{figure}

We now study the phase transition along the cut \cE~ in detail. We start with studying the finite-size behavior of the Binder cumulants $U_m$ and $U_\phi$  at 
cut \cE. The numerical results of $U_m$ and $U_\phi$ as functions 
of $g_\mathcal{E} =Q_3/J$ for several system sizes are shown in Fig.\ref{E-binder}. For $g_\mathcal{E}<1.63$, $U_m$ converges to 1 as system
size increases, suggesting a N\'eel ordered state. For $g_\mathcal{E}>1.63$, $U_\phi$ converges to 1 as system
size increases, suggesting a VBS-ordered state. A direct transition between VBS and N\'eel states is evident. 
Both of the two Binder cumulants tend  negative near the transition point as system size increases, indicating that this phase transition is possibly first order \cite{Binder-firstorder}.

\begin{figure}[htbp]
\centering
\setlength{\abovecaptionskip}{20pt}
\includegraphics[angle=0,width=0.4\textwidth]{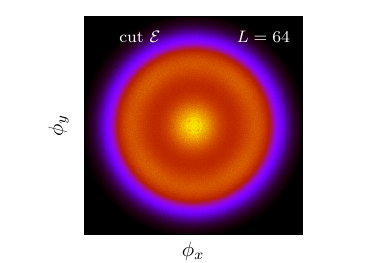}
\caption{Histogram of the VBS order parameter $\vec{\phi}$ of one layer for $L=64$ at  $Q_3/J=1.639$ near cut $\mathcal{E}$. The brighter color indicates a higher probability of occurrence, with the $x$ and $y$ axes representing $\phi_x$ and $\phi_y$ ranging from $-0.05 \sim 0.05$. 
}
\label{EG-his}
\end{figure}

To further investigate the properties of the transition, we study the histogram of $\vec \phi$ at $Q_3/J=1.639$ near cut \cE, the results are presented in Fig.\ref{EG-his}. There are 
bright peaks at the origin and at a ring with a finite radius, indicating the coexistence of the N\'eel phase and the VBS phase, thereby confirming that the transition is first-order.  
Compared to the first-order N\'eel-VBS transition for the S-S coupled model at cut \cB ~and \cC, much larger system sizes are required to observe the coexistence of two states in the histogram
of $\vec{\phi}$
for the E-E coupled bilayer $J$-$Q_3$ model.
This indicates that the first-order phase transition in the E-E coupling model is weaker than that in the S-S coupling model.
More interestingly, the histogram peak associated with the VBS state shows no angular dependence at the coexisting point, suggesting that the transition might be close to a critical point with
\reply{emergent U(1) symmetry}.

\section{Summary}\label{sec:sum}

In this paper, we have studied the phase diagrams and phase transitions of models with bilayers of $S=1/2$ square lattice antiferromagnets with SU(2) Heisenberg symmetric interactions constructed from 
singlet projector on pairs of spins, creating two-spin,  four-spin and six-spin 
interactions. The models we studied can be classified into two main categories based on the interlayer 
interactions, which exhibit different internal symmetries, thus resulting in distinct phase diagram topologies. 
The first category is the conventional bilayer models, where the interlayer interactions are of the Heisenberg type, 
allowing the two layers to exchange spins and energy, leading to a simple disordered dimer state. The resulting phase diagram is quite rich, including N\'eel and VBS phases along with dimer phases, and features both 
first-order and continuous phase transitions. In the second category of models, the layers can only exchange energy but not spin, preventing
the formation of the trivial dimer phase; thus, the phase diagram includes only the N\'eel and the VBS phases. We find evidence for first-order behavior and coexistence of N\'eel and VBS phases. Surprisingly, however, in the coexistence region the VBS order histogram for 
\cE~ does not show the expected four-fold anisotropy (as seen for example in \cB) but instead displays a U(1) symmetry. This is unexpected and is indicative of an emergent symmetry or a proximity to a continuous transition.

\reply{Regarding experimental relevance, we note that the core physics of our model, the competition between Néel and VBS orders, has been a focal point in recent experimental studies. A prominent example is the Shastry-Sutherland compound SrCu$_2$(BO$_3$)$_2$, where high-pressure experiments have revealed a transition between a plaquette-singlet state and an antiferromagnetic state. \cite{Zayed2017,Cui2023}}

Our study of the phase diagram of bilayer models uncovers a phase transition for which we do not yet have a full understanding and which merits further study. In the S-S coupled bilayer, we found that the phase transition at the cut $\mathcal{D}$ (between the VBS and dimer phases) is continuous but one in which
the $Z_4$ anisotropy survives even on the largest lattices. This is inconsistent with the simplest expectation for the critical phenomena, i.e. the XY model with a dangerously irrelevant four-fold anisotropy scenario. An interesting origin for this difference could be that in the VBS order parameter the anisotropy is locked to the lattice and not simply to an internal order parameter as it is in the usual XY model with four-fold anisotropy. This issue was considered in a classical two-dimensional model in Ref.\cite{gCM}. An extension of this study to 2+1 dimension, which is beyond the scope of the current work, may be a promising way to understand this issue further, we leave this for future investigation.

\begin{acknowledgments}
We thank A.W. Sandvik for helpful discussions.
This work was supported by the National Natural Science Foundation of China under Grant No. 12175015 (FZ, WG). \reply{RKK was supported in part by the NSF Award No. DMR-2312742.}
\end{acknowledgments}

\section*{Data Availability}
The data that support the findings of this study are available in Ref.~\cite{dataset}.

%

\end{document}